\documentstyle[aps,epsfig,rotating]{revtex}

\def\bild#1#2{\epsfig{file={#1},width=#2}}

\def\beq{\begin{equation}}
\def\eeq{\end{equation}}

\def\reff#1{(\ref{#1})}

\def\vekt#1{\bbox{#1}}

\def\vektr{\vekt{r}}

\def\vektnabla{\vekt{\nabla}}

\def\operator#1{\mbox{\sf #1}}

\def\Nrem{N_{\mbox{\scriptsize rem}}}
\def\tend{t_{\mbox{\scriptsize end}}}

\def\Vop{{\operator{V}}}

\def\halb{\frac{1}{2}}

\def\Adach{\hat{A}}

\def\Vee{V_{\mbox{\rm\scriptsize ee}}}
\def\uxcisigma{{u_{\mbox{\rm\scriptsize xc}}}_{i\sigma}}
\def\Vxc{V_{\mbox{\rm\scriptsize xc}}}
\def\Vxcsigma{{\Vxc}_\sigma}

\def\Vxcslatersigma{V_{\mbox{\rm\scriptsize xc} \sigma}^{\mbox{\rm\scriptsize Slater}}}

\def\Vxcsigmaxlda{{V_{\mbox{\rm\scriptsize xc}}^{\mbox{\rm\scriptsize XLDA}}}_{\!\!\!\!\!\!\!\!\!\!\!\!\sigma\,\,\,\,\,\,\,\,\,}}

\def\pabl#1#2{\frac{\partial #1}{\partial #2}}

\def\imagi{\mbox{\rm i}}
\def\diff{\,\mbox{\rm d}}

\begin{document}

\title{C$_{60}$ in intense femtosecond laser pulses: nonlinear dipole response and ionization}
\author{D.~Bauer,$^1$ F.~Ceccherini,$^{1,2}$ A.\ Macchi,$^2$ and F.~Cornolti$^2$}
\address{$^1$Theoretical Quantum Electronics (TQE), Darmstadt University of Technology, Hochschulstr.~4A, D-64289 Darmstadt, Germany\\ $^2$INFM, sez.\ A, Dipartimento di Fisica, Universit\`a di Pisa, Piazza Torricelli
    2, 56100 Pisa, Italy}

\date{\today}

\maketitle

\begin{abstract}
We study the interaction of strong femtosecond laser pulses with the C$_{60}$ molecule 
employing time-dependent density functional theory with the ionic background treated in a jellium approximation.
The laser intensities considered are below the 
threshold of strong fragmentation but too high for perturbative treatments such as linear
response.
The nonlinear response of the model to excitations by short pulses of frequencies up to 45\,eV is presented and analyzed 
with the help of Kohn-Sham orbital resolved dipole spectra. 
In femtosecond laser pulses of 800\,nm wavelength  ionization is found to occur multiphoton-like rather than via excitation of a ``giant'' resonance. 

\bigskip
\noindent PACS numbers: 36.40.-c, 33.80.Rv, 31.15.Ew

\end{abstract}

\pacs{36.40.-c, 33.80.Rv, 31.15.Ew}

\section{Introduction}
Intense laser atom interaction exhibits nonlinear phenomena such as above threshold ionization (ATI), high harmonic generation (HHG), and nonsequential multiple ionization (NSI) (see \cite{revs} for recent reviews). While some of those features are accessible  in terms of a sequential ``single active electron'' (SAE) approach others are clear manifestations of many electron effects and correlation, e.g., NSI. The full {\em ab initio}\ solution of the time-dependent Schr\"odinger equation (TDSE) for two active electrons interacting in their full dimensionality with the laser and their parent ion is already at the limit of what is possible with modern computers \cite{taylor}. Treating many electron systems in laser fields thus needs further approximations. Density functional theory (DFT), extremely successful in electronic structure calculations  of many-electron systems ( see, e.g., \cite{dreizgross}), has been extended to the time-dependent case (TDDFT) (see \cite{runge}, and, e.g., \cite{burke} for a review). Despite the fact that TDDFT still lacks an equally solid foundation compared to that on which ground state DFT is built, it was successfully applied to metal clusters in laser pulses (see \cite{metclus} for a review). Problems mainly arise when observables have to be extracted which are not obvious functionals of the Kohn-Sham orbitals or the total electron density, like in the study of NSI of atoms within TDDFT \cite{tddftnsi,baueroe}, or when the results are very sensitive to the choice of the approximation to the unknown exchange-correlation potential. 
Compared to laser atom interaction, in big molecules or clusters additional degrees of freedom are introduced: electronic degrees of freedom, including collective effects such as the formation of plasmons, vibrational degrees of freedom, or fragmentation. With laser pulses of different duration the equilibration of energy among the various channels can be probed. For C$_{60}$ this was nicely demonstrated in Ref.~\cite{campb} where the photoelectron spectra in fs laser pulses exhibited ATI peaks, a signature for direct multiphoton processes, which disappeared for longer pulses where collective effects set in. Concerning the ionization mechanism of C$_{60}$ in fs laser pulses there is a discrepancy in the literature. While in the recent work of Tchaplyguine {\em et al.}\ \cite{tchap} from ion yield-curves vs.\ laser intensity direct multiphoton ionization was found to be the responsible pathway for ionization of C$_{60}$, in an earlier publication Hunsche {\em et al.}\ \cite{hunsche} claimed it is the excitation of a giant resonance near 20\,eV. Such a  resonance at 20\,eV in C$_{60}$ was first predicted theoretically by Bertsch {\em et al.}\ \cite{bertsch} and confirmed later in an experiment by Hertel {\em et al.}\ \cite{hertel} using synchrotron radiation for single-photon ionization measurements. When compared to metal clusters where collective resonances occur at a few eV a 20\,eV giant resonance of about 12\,eV width is quite remarkable.

The nonlinear TDDFT treatment of C$_{60}$ in a laser pulse is numerically rather demanding because one has to allow for ionization which implies the use of a big numerical grid in order to represent the time-dependent Kohn-Sham (KS) orbitals properly. It is thus impossible, at least with computers nowadays available,  to achieve both a detailed account of the soccer ball-like structure of C$_{60}$  {\em and}\ an accurate propagation of freed electron density far away from the ion and, possibly, back. Therefore we restrict ourselves to a jellium approach for the ionic background of the C$_{60}$. Such a jellium model was employed in \cite{puska} to study the photo absorption of atoms inside C$_{60}$ within linear response theory. It was found to share many of the relevant features with more demanding ``first principle'' calculations (like, e.g., in  \cite{bertsch}) and experiment \cite{hertel}. Jellium models were also successfully applied to metal clusters (see \cite{brack} for a review).

The paper is organized as follows. In Section \ref{model} we present our TDDFT jellium model of laser C$_{60}$ interaction. In Section \ref{dipresp} we characterize the dipole spectrum of our model in terms of single particle transitions. In Section \ref{ionpla} we study the dipole response and ionization after excitation by short pulses of different frequencies. In Section \ref{sloperes} we examine the ionization mechanism of our C$_{60}$ model in 800\,nm fs laser pulses. Section \ref{concl} contains a brief summary and conclusion.

\section{The model: static properties} \label{model}
The time-dependent Kohn-Sham (TDKS) equation for the orbital $\Psi_{i\sigma}(\vektr,t)$ reads (atomic units are used unless noted otherwise)
\beq \imagi \pabl{}{t} \Psi_{i\sigma}(\vektr,t) = \left( -\halb \vektnabla^2 + V(r) + \Vop_I(t) + {\Vee}_\sigma[n_\uparrow(\vektr,t),n_\downarrow(\vektr,t)] \right)   \Psi_{i\sigma}(\vektr,t) . \eeq
Here, $\sigma=\uparrow,\downarrow$ indicates the spin polarization,  $V(r)$ is the potential of the ions, $\Vop_I(t)$ is the laser in dipole approximation, and ${\Vee}_\sigma [n_\uparrow,n_\downarrow]$ is the effective electron-electron interaction potential which is a functional of the electron spin densities 
\beq n_\sigma(\vektr,t) = \sum_{i=1}^{N_\sigma}  \vert\Psi_{i\sigma}(\vektr,t)\vert^2.\eeq
$N_\sigma$ is the number of orbitals occupied by KS particles with spin $\sigma$. The total electron density is 
\beq n(\vektr,t)=\sum_\sigma n_\sigma(\vektr,t).\eeq  
The electron-electron potential is splitted, 
\beq {\Vee}_\sigma[n_\uparrow,n_\downarrow] = U[n] + \Vxcsigma[n_\uparrow,n_\downarrow], \eeq
where $U[n]$ is the Hartree potential
\beq U[n]= \int\!\! \diff^3 r'\, \frac{n(\vektr',t)}{\vert\vektr-\vektr'\vert}\label{hartree} \eeq
and $\Vxcsigma[n_\uparrow,n_\downarrow]$ is the exchange correlation (xc)-part.  Although the Runge-Gross theorem \cite{runge} ensures that, in principle, the time-dependent KS scheme could yield the exact density $n(\vektr,t)$ on which all observables depend, in practice an approximation to the exchange-correlation potential $\Vxcsigma[n_\uparrow,n_\downarrow]$ has to be made. 
We chose the Slater expression 
\beq \Vxcslatersigma(\vektr,t)=\sum_{i=1}^{N_\sigma} \frac{n_{i\sigma}(\vektr,t)}{n_\sigma (\vektr,t) } \uxcisigma(\vektr,t) ,  \label{slater} \eeq
where $\uxcisigma=\Vxcsigmaxlda [n_\uparrow,n_\downarrow] - U[n_{i\sigma}] - \Vxcsigmaxlda [n_{i\sigma},0]$, i.e., the self-interaction is removed, and the exchange-only local density approximation (XLDA) was employed. In TDDFT simulations of nonsequential ionization in laser atom interaction \cite{baueroe} we found that, compared to plain XLDA, the Slater potential \reff{slater} improved the ionization potentials and the fulfillment of Koopman's theorem significantly. Instead, for the kind of studies presented in this paper it is actually not necessary to go beyond XLDA. Inaccuracies in the, e.g., ionization potential for the outermost electron can be easily compensated by adjusting the free parameters of the jellium potential for the ionic background (see below). In Ref.\ \cite{ullr}, for sodium clusters in laser pulses a self interaction correction was found to cause ``minor differences'' in dipole spectra  leaving  the ionization mechanism ``essentially unchanged.''    
Adding a correlation potential (we used the one proposed by Perdew and Wang \cite{perdwang}) had negligible effects on the results presented in this work. 

Usually the dipole approximation is said to be justified when the system under study is small with respect to the laser wavelength. However, when a light beam impinges on an object smaller than the wavelength the interior could be, nevertheless, field-free,  provided the electron density is sufficiently high. This kind of screening would not be included in a TDDFT treatment with dipole approximation.  The condition for laser light of wavelength $\lambda_i$ and frequency $\omega_i$ being able to penetrate a plasma layer of thickness $\delta$ is $\xi=\pi (\omega_p/\omega_i)^2 \delta/\lambda_i<1$ \cite{vshiv}. This is the case for our parameters. Therefore, the dipole approximation is safe also in this respect.

In our numerical code, the KS orbitals are expanded in spherical harmonics $Y_\ell^m(\vartheta,\varphi)$. If the ground state has a closed shell structure the effective KS potential is spherical. Hence $\ell$ and $m$ are ``good'' quantum numbers for the ground state configuration. In a linearly polarized laser field (in dipole approximation) the quantum numbers $m$ remain good, i.e., there is no $m$-coupling through the laser because the azimuthal symmetry is retained.  $\Vop_I(t)$ introduces an $\ell\pm 1$-coupling only. However, orbitals with different $\vert m\vert$ behave differently in the laser field. The radial KS wave functions are discretized in position space. Each time step, the effective potential has to be calculated which makes a TDKS solver significantly more time-consuming than an ordinary TDSE code running a corresponding SAE problem. The effective potential $\Vee$  was expanded up to the dipole. Consequently, both the laser and $\Vee$ lead to an  $\ell\pm 1$-coupling only. We are confident that neglecting higher order multipoles of $\Vee$ does not affect the validity of our conclusions \cite{explaindipole}.
The actual propagation is performed in velocity gauge using an algorithm similar to the one for the TDSE described by Muller in Ref.~\cite{muller}. Probability density reaching the radial boundary of the numerical grid at $\approx 100$\,a.u.\ was removed by an imaginary potential. The eventually decreasing norm then can be interpreted as one minus the ionization probability of that orbital.

The laser is polarized in $z$-direction so that in velocity gauge we have
\beq \Vop_I(t)=-\imagi A(t) \pabl{}{z} \eeq
where $A(t)$ is the vector potential and the $A^2$-term has been transformed away (see, e.g., \cite{faisalbook}).
This potential leads to the above mentioned $\ell\pm 1$ coupling.

The ionic background is treated in a jellium approximation, i.e., the ions are thought of being smeared out over a spherical shell with outer and inner radius $R_o$ and $R_i$, respectively. The ionic charge density is constant for $R_o>r>R_i$ and zero otherwise. The radii $R_o$, $R_i$ are centered around the known radius of the C$_{60}$-fullerene, $(R_o+R_i)/2=R=6.7$\,a.u. In real C$_{60}$ there are 60 $\pi$-electrons and 180 $\sigma$-electrons. Therefore the charge of the jellium background should be 240\,a.u. However, 240 KS particles do not yield a self-consistent closed shell-structure for the ground state of our model. Since partially filled shells would spoil the spherical symmetry of the ground state  we take 250 electrons instead (see also \cite{puska}) which leads to a self-consistent closed shell ground state of the jellium model.
Thus, introducing 
\beq r_s^{-3}=\frac{N}{R_o^3-R_i^3}, \qquad N=250, \eeq
and allowing for an additional potential depth parameter $v_0$ we arrive at an ionic potential
\[ V(r)= \left\{ \matrix{ -r_s^{-3} 3 (R_o^2-R_i^2)/2 & \mbox{\rm for} &  r\leq R_i, \cr 
 -r_s^{-3} \left( 3 R_o^2/2 -\left[  r^2/2 + R_i^3/r \right] \right) - v_0 &\mbox{\rm for} &  R_i<r<R_o, \cr
 -r_s^{-3} (R_o^3-R_i^3)/r &\mbox{\rm for}&  r\geq R_o . } \right. \]
The parameters $R_o$, $R_i$, and $v_0$ can be varied in order to obtain a jellium-C$_{60}$ ground state which shares the relevant features with first-principle calculations of ``real'' C$_{60}$ (see also Ref.\,\cite{puska}). E.g., by decreasing $R_o-R_i$ the overlap of the $\sigma$ and the $\pi$ single particle levels decreases. With $v_0$ the absolute position of the level scheme can be adjusted in order to meet, e.g., the ionization potential for the outermost electron. We chose  $R_o=8.1$, $R_i=5.3$, and $v_0=0.78$. Some of the ground state properties of the model are shown in Fig.~\ref{potpic}. Because of the centrifugal barrier $\ell(\ell+1)/2r^2$ the total potential is $\ell$-dependent, and states with high $\ell$ are pushed outwards. The energy levels are $2(2\ell+1)$-degenerated. The 250 KS particles can be subdivided in 200 $n=1$-states (the $\sigma$-electrons) occupying $\ell$-values from $0$ up to $9$, and $50$ $n=2$-states (the $\pi$-electrons). There are also bound $n=3$-states ($\delta$-levels) but they are not occupied in the ground state configuration.  The orbital densities are also shown in Fig.~\ref{potpic}. Each $\pi$-electron wavefunction has a node near the jellium-shell radius $R$. The values of the single KS particle orbital eigenenergies are given in Table~\ref{tabnl}. The highest occupied state is the $\pi$ state with $\ell=4$. From Koopman's theorem we therefore expect an ionization energy of $I_p^+=0.274$. Calculating the ionization energy by subtracting the total energy of C$_{60}^+$ from that for neutral C$_{60}$ we obtain (on our numerical grid with a grid spacing $\Delta r=0.3$, and the parameters $R_i,R_o,v_0$ as chosen above) $I_p^+=0.279$ (7.59\,eV) which agrees reasonably well with the former value and experiment ($\approx 7.6$\,eV, \cite{seifert}). In any case we expect for 800\,nm laser light five photons being necessary for removing the outer electron. 
However, collective effects might occur so that more photons are required. In fact, there is an unresolved discrepancy in the literature about whether ionization of C$_{60}$ in the fs-regime works multiphoton-like \cite{tchap} or through the excitation of a 20\,eV giant resonance \cite{hunsche}. The results from our model concerning this question will be presented in Section~\ref{sloperes}.

In Table~\ref{tabnl} we also enumerated the KS orbitals for the sake of easy reference later on. Since in each $\ell$-shell there are $\ell+1$ different $\vert m\vert$-values we need 70 KS orbitals to describe our jellium-model interacting with the laser field. In each $\ell$-shell the KS orbitals are labeled from $m=0$ up to $\vert m\vert=\ell$. Thus, e.g., orbital no.\ 0 refers to the two electrons of opposite spin which, in the ground state configuration, populate the $n=1$, $\ell=m=0$ $\sigma$-state whereas orbital no.\ 69 is initially a pure $n=2$, $\ell=\vert m\vert=4$ $\pi$-state, populated by four electrons with different spin and/or sign of $m$.

\section{Dipole response} \label{dipresp}
Within linear response theory excitations and giant resonances are commonly inferred from the photo absorption cross section, i.e., the imaginary part of the polarizability. Here we follow the different route of nonlinear TDDFT which allows us to distinguish between ionization, single particle transitions, or plasmons, and also accounts for higher order processes beyond single particle hole excitations.

First we discuss the nonlinear response of our C$_{60}$ model to a kick at time $t=0$ by a delta-like electric field $E(t)=\Adach\delta(t)$. Such a kick is equivalent to the method proposed by Yabana and Bertsch in Ref.\,\cite{yabana} where the ground state wavefunctions are perturbed by giving them a coherent velocity field, i.e., $\Psi_i\to\exp(\imagi k z) \Psi_i$, in order to initiate a dipole moment evolving in time afterwards. From the Fourier-transform of the dipole $d(t)=\int z n(\vektr)\, \diff^3 r$ the frequency dependent response is obtained. In Fig.\,\ref{kickksord} the Kohn-Sham orbital resolved dipole (KSORD) response for a delta-kick of magnitude $\Adach=0.01$\,a.u.\ is presented. Such a kick can still be considered a small perturbation to the ground state configuration because only a fraction $3.2\times10^{-5}$ of the total electron density was freed (corresponding to an ionization probability of $8\%$ for the outermost electron). On the left a contour plot of the dipole vs.\ orbital number and time is shown. On the right-hand-side the corresponding spectrum is plotted, obtained by Fourier-transforming the dipoles of the individual KS orbitals. The total dipole and its Fourier transform are also included in Fig.\,\ref{kickksord}. Looking at the dipoles vs.\ time one easily identifies the different $\ell$-shells. The $\sigma$-electrons are KS orbital numbers 0--54, the $\pi$-electrons range from 55--69. Examining the KSORD spectra on the right one clearly identifies rather narrow vertical lines. Each of those vertical lines can be understood as a single particle transition between ground state KS levels. For a certain KS orbital the dipole strength is particularly high for those transitions where one of the two levels involved is the one which is occupied in the ground state configuration of this KS orbital. This explains the parabola-like structures of strong dipole emission visible in the KSORD spectra. 

The low frequency lines in the range 0--5\,eV stem from transitions of the type $\ell\to\ell\pm 1$ with the $n$-quantum number fixed. Lines between 5 and 15\,eV are caused by transitions of type $\pi\ell\to\delta(\ell\pm 1)$, and are therefore particularly pronounced for orbitals representing $\pi$-electrons initially. The lines with high dipole strength  along a parabola-like structure around 20\,eV for both, $\pi$- and $\sigma$-electrons originate from transitions of the type $\sigma\ell\to\pi(\ell+1)$ or $\pi\ell\to\sigma(\ell-1)$ (right branches) and $\sigma\ell\to\pi(\ell-1)$ or $\pi\ell\to\sigma(\ell+1)$ (left branches). 
Even higher dipole emission around 30\,eV stems from $\sigma\delta$-transitions.  A common feature for all orbitals where $\ell=\vert m\vert$ initially (i.e., orbitals number 0, 2, 5, 9, 14, 20, 27, 35, 44, 54, 55, 57, 60, 64, 69) is the relatively strong emission around 30\,eV. The reason why the $\ell=\vert m\vert$-orbitals preferably radiate at those frequencies is that $\ell\to\ell-1$-transitions are not possible for them. Indeed, the left branch of the parabola-like structure for the $\sigma$-electrons (related to transitions where the $\ell$ quantum number decreases by one) shows gaps for the $\ell=\vert m\vert$ orbitals whereas for the $\pi$ electrons it is the right branch where the corresponding lines are missing. 
The spectrum of the total dipole is shown above the contour plot. It is a broad structure fragmented in lines originating from single particle transitions. Around 20\,eV there is a minimum in the total spectrum although the KSORD spectra show lines there. This is due to destructive interference. Instead, the dipole emission from KS orbitals with (initially) high $\ell$ quantum number at 17 and 24\,eV interferes constructively, leading to two pronounced peaks in the total dipole spectrum. The small zero frequency component is caused by the few electron density which was freed by the delta-kick. 

The dipole spectrum in Fig.\,\ref{kickksord} can be compared with the linear response results in \cite{puska} (Fig.\,3)  and in \cite{alasia} (Fig.\,1). In Ref.\,\cite{puska} where a similar jellium model was studied within linear response theory we find similarities with our spectrum in the energy region 14--20\,eV, in particular the pronounced 17\,eV-line. Instead, in Fig.\,3 of \cite{puska} there is weak dipole emission below 10\,eV (apart from a peak at 4\,eV) and also low dipole strength above 25\,eV, apart from a broad but rather weak hump with a maximum around 34\,eV. In our spectrum in Fig.\,\ref{kickksord} lines in the region 25--35\,eV are much stronger  because also transitions which are not directly accessible by single particle processes starting from the ground state configuration are allowed in the nonlinear treatment. Similar to our result, in  Fig.\,1a of Ref.\,\cite{alasia} a highly fragmented structure is visible in the particle-hole strength function for energies 4--15\,eV.

\section{Excitation and ionization} \label{ionpla}
We calculated the interaction of our jellium model with ten cycle sin$^2$-shaped (with respect to the electric field) laser pulses in the frequency range from 6.8  up to 47.6\,eV. The peak intensity was adjusted in such a way that intensity times pulse duration (energy per unit area) was held constant, i.e., in atomic units $\Adach^2\omega_i = 1.8375\times 10^{-5}$ where $\omega_i$ is the incident laser frequency and $\Adach$ is the vector potential amplitude. With such laser intensities the probability to remove the first electron remained below 10\% for all frequencies. After the laser pulse was over we continued the propagation of all KS orbitals to allow for delayed ionization and free oscillations. The total simulation time corresponded to $\tend=11.5$\,fs so that a real C$_{60}$ which has been  ionized at most to C$^+_{60}$ has no time for fragmentation. 
The result is presented in Fig.~\ref{spekafterone}. The contour plot shows the logarithmically scaled dipole strength vs.\ incident frequency $\omega_i$ and emitted frequency $\omega_e$. Left to the contour plot the number of removed electrons
\[ \Nrem=250-\int_{\mbox{\rm\scriptsize Grid}}\diff^3r\, n(\vektr,\tend) \]
is plotted as a function of the incident frequency $\omega_i$. In all cases delayed ionization was negligible compared to the electron density which was freed owing to the laser pulse.

The maximum dipole strength is along  the diagonal $\omega_i=\omega_e$ where the excitation was resonant. However, at $\omega_i=\omega_e=20.5$\,eV there is a relatively weak dipole response but the ionization yield has a local maximum (dashed line). Instead, the local minima in ionization at 17 and 24\,eV coincide with emission in two strong lines near the same values for $\omega_e$. Those lines were already observed in Section \ref{dipresp}. 
For $\omega_i=30$\,eV one observes both, strong ionization {\em and} a strong dipole response. For even higher $\omega_i$ ionization drops and the jellium system does not provide modes $>40$\,eV.

The mutually exclusive behavior of ionization and dipole emission at 17, 20.5, and 24\,eV might be surprising. One could expect that resonantly driven transitions yield big excursions of the electron density, thus leading to strong ionization also. This was observed for the case of sodium clusters interacting with strong laser pulses of 100\,fs duration \cite{ullrii}, and also the experimental results in \cite{hertel} were analyzed by unfolding excitation and (subsequent) autoionization. We believe that our different findings are due to the short duration of the exciting laser pulse.  Although at 17 and 24\,eV the short laser pulse excites resonantly $\pi\sigma$-transitions with the outermost electron involved it does not efficiently couple those excitations to the continuum, thus keeping the ionization probability relatively low. Instead, at 20.5\,eV the outer electron is not resonantly coupled to another {\em bound} state but rather absorbs a single photon to make a bound-free transition. The transitions that are resonant with the incident frequency of 20.5\,eV involve low-$\ell$ $\sigma$- and $\pi$-orbitals whose emitted radiation interfere destructively, as discussed already in Section \ref{dipresp}. At 30\,eV the outer electron can be freed by absorbing a single photon while stronger bound electrons are resonantly excited to perform bound-bound transitions of $\sigma\delta$-type which leads to efficient dipole emission also. Concluding this Section we can say that in very short exciting laser pulses the C$_{60}$ response seems dominated (i) by single particle transitions, although not exclusively those starting from the ground state, and (ii) by {\em direct} photoionization instead of ionization by resonantly driven single particle transitions or plasmons, as it is the case in longer pulses.

\section{Ionization mechanism at $\lambda=800$ nanometer}\label{sloperes} 
After studying the dipole response at relatively high energies we now turn to the interaction of C$_{60}$ with laser light of 800\,nm wavelength. 
We simulated the interaction of our C$_{60}$ jellium model with a ten cycle 800\,nm sin$^2$-shaped laser pulses (corresponding to 26\,fs pulse duration).
In Fig.~\ref{slope} the removed electron density $\Nrem$ after the pulse vs.\ the peak intensity of the pulse is presented. In the regime where $\Nrem\ll 1$ this is equivalent to the single ionization probability. In the case of perturbative off-resonant multiphoton ionization of atoms it is well-known (see, e.g., \cite{faisalbook}) that the ionization probability is $\sim I^n$ where $n$ is the number of photons needed to reach the continuum from the initial state. From Section \ref{model} we know that in our jellium model the first ionization potential is $I_p^+=0.279$, and thus we expect $n=5$ photons necessary for ionization if C$_{60}$ behaves multiphoton-like. As is evident from Fig.~\ref{slope} this is the case, in agreement with  an experiment performed using a Ti Sapphire laser ($\lambda=800$\,nm, 30 to 120\,fs pulse duration) \cite{tchap}. There, from the C$_{60}^{2+}$ yield-slope, also the second electron was found to behave multiphoton-like with $n=8$. Those findings are in contrast with earlier experimental results in \cite{hunsche} (same wavelength and pulse durations) where excitation of the 20\,eV Mie-resonance, corresponding to $n=13$ photons, was concluded to be the dominant ionization mechanism. The $n=13$ slope is depicted in  Fig.~\ref{slope} in the upper left corner and is much too steep to fit with our numerical result. This appears reasonable because the incident laser frequency 1.6\,eV lies energetically far below 20\,eV so that in short pulses an efficient excitation of the latter is unlikely. The experimental fragmentation onset and the C$_{60}$ saturation intensity as observed in \cite{tchap} is also indicated in Fig.~\ref{slope}.

In Fig.~\ref{panelafterpulse} we present dipoles and the corresponding spectra for two particular laser intensities. For the higher intensity [plots (c) and (d)] the electron density continues oscillating quite strongly after the pulse but with little effect on ionization.  In the spectra (b) and (d) the black curve was calculated from Fourier-transforming the dipole with respect to the entire time interval shown in the plots above. Therefore lines corresponding to the laser harmonics can be observed in those spectra. Fourier-transforming only over the time after the pulse leads to the gray curves. In the low intensity case (b) the dominant line is around 3\,eV, corresponding to the transition of the outermost $\pi$-electron from $\ell=4$ to $\ell=5$. This transition is nearly resonant with two laser photons. In the high intensity case the electron density continues oscillating with the laser frequency even after the pulse. The next more energetic peaks are, again, the $\ell=4\to 5$ transition near 3\,eV followed by the $\ell=9\to 10$ transition of a $\sigma$-electron at 5.5\,eV. The next line corresponds to the ionization energy 7.5\,eV. For both intensities excitations beyond 10\,eV are several orders of magnitude weaker and can therefore not play any role in the ionization process of our C$_{60}$ model at that laser wavelength.

\section{Conclusion} \label{concl}
In this paper we presented a nonlinear time-dependent density functional theory (TDDFT)-treatment of a  C$_{60}$ jellium model in intense laser pulses. The Kohn-Sham (KS) orbitals were expanded in spherical harmonics and discretized in radial direction on a sufficiently big numerical grid on which they were propagated in time. By this method all bound states and the continuum can be properly represented, allowing for an accurate description of ionization and higher order transitions, not included in linear response theory (sometimes also called TDDFT).

We characterized the KS orbital resolved dipole (KSORD) response. The KSORD spectra can be understood in terms of single particle transitions. The dipole radiation emitted by individual KS orbitals can interfere destructively. This is the case for $\sigma\pi$-transitions  between KS orbitals with low $\ell$ (and $\ell\pm 1$). The dominant lines in the total dipole spectrum at 17 and 24\,eV were found to originate from single particle $\sigma\pi$-transitions with high-$\ell$ KS orbitals involved. We also found a strong response in the dipole spectra around 30\,eV. This broad structure is fragmented in single particle transitions of $\sigma\delta$-type.

For short exciting pulses with frequencies between 7 and 48\,eV we studied both ionization and the dipole spectra, calculated from the nonlinear dipole response after the pulse had passed by. We observed that when the incident frequency matched with the two resonant peaks of the dipole spectrum at 17 and 24\,eV, respectively, ionization dropped. Instead, an incident frequency in between ($\approx 20.5$\,eV) led to a maximum ion yield. This was attributed to the fact that in very short exciting laser pulses the C$_{60}$ response is dominated by direct photoionization instead of by ionization due to resonantly driven single particle transitions or plasmons. Hence, when the outermost electron is coupled resonantly to another {\em bound} state ionization decreases. A maximum in both ionization yield and dipole strength was found at frequencies around 31\,eV where the outer $\pi$ electron is directly photoionized and stronger bound electrons are resonantly excited. For even higher incident photon energies $>37$\,eV ionization decreased because energy transfer to the C$_{60}$  became inefficient.

For 800\,nm fs laser pulses we observed direct multiphoton ionization rather than ionization via excitation of a ``Mie-resonance''. This result agrees with recent experimental findings in \cite{tchap}. Although  it disagrees with earlier results in \cite{hunsche} we are quite confident that our model yields correct predictions  in this respect because the multiphoton character for optical (or near optical) frequencies is well-known for metal clusters (see, e.g., \cite{metclus}, and references therein) where collective electron behavior is more pronounced and resonances are less energetic.

The TDDFT jellium model offers one of the very few feasible approaches for theoretical investigations concerning  the interaction of intense laser light with complex systems where many electron-effects, bound-free transitions, and rescattering might be important. Preliminary results for ``above threshold ionization'' (ATI) of C$_{60}$ (as in the experiment \cite{campb}) were also obtained within this model \cite{bauerati}. Simulating ATI is always much more demanding than studying the dipole response and ionization only since one has to trace the free electron motion in the continuum as well as rescattering events with very high accuracy over several laser cycles of optical frequency.

\section*{Acknowledgment}
This work was supported by the FUMOFIL project through the INFM Parallel Computing Initiative, and by the Deutsche Forschungsgemeinschaft within the SPP ``Wechselwirkung intensiver Laserfelder mit Materie.'' Illuminating discussions with D.\ Pitrelli are gratefully acknowledged.

\pagebreak

\begin{table}
\caption{\label{tabnl} The ground state single Kohn-Sham (KS) particle orbital energies as obtained on our numerical grid used for propagation (grid spacing $\Delta r=0.3$). The $\sigma$-electrons occupy $\ell$-values up to 9, the $\pi$-electrons up to 4. The $\delta$-orbitals are not occupied in the ground state configuration. Orbital energies of unoccupied levels are written {\em italic}. The occupied orbitals are enumerated for the sake of easy reference. For each $\ell$-shell $\ell+1$ KS orbitals to account  for all the possible $\vert m\vert$ (numbered from lowest to highest) are needed. }
\end{table}

\begin{figure}
\caption{\label{potpic}} Ground state poperties of the jellium C$_{60}$-model. (a) The total potential $V(r)+U[n(r)]+\Vxc[n(r)]$ depends on the angular quantum number $\ell$ (through the centrifugal barrier). For $n=1$ ($\sigma$-electrons) orbitals from $\ell=0$ up to $\ell=9$ are occupied in the ground state situation, for $n=2$ ($\pi$-electrons) orbitals from $\ell=0$ to $\ell=4$ are occupied. The potential for the empty $\ell=10$ orbitals is drawn dashed. The radial shape of the total ground state density $n(r)$ is also plotted. (b) The single Kohn-Sham particle energy levels corresponding to the potentials in (a). $\sigma$-states are drawn solid while $\pi$-states are plotted dashed. The degeneracy is $2(2\ell +1)$. (c) The orbital densities $n^{(\sigma,\pi)}_\ell$. The sum of those is $n(r)/4\pi$. 
\end{figure}

\begin{figure}
\caption{\label{kickksord}} Kohn-Sham (KS) orbital resolved dipoles (KSORD, left) and the corresponding spectrum (right) after a delta-kick with an electric field $E(t)=\Adach\delta(t)$, $\Adach=0.01$. The Kohn-Sham orbitals are enumerated according Table~\ref{tabnl}. The total dipole $d(t)$ and the total dipole strength (Dip.str) are given also. See text for a detailed discussion.
\end{figure}

\begin{figure}
\caption{\label{spekafterone}} Right-hand-side: contour plot of the Fourier-transformed dipole $d(t)$ after ten cycle pulses of frequency $\omega_i$. The dipole strength vs.\ incident frequency $\omega_i$ and emitted frequency $\omega_e$ is logarithmically scaled (cf.\ color bar ranging from $10^{-30}$ to $10^{-7}$, in arbitrary units). Left-hand-side: number of removed electrons. The horizontal solid (dashed) lines (line) indicate frequencies $\omega_i$ where ionization was relatively low (high) and excitation was efficient (inefficient).   
\end{figure}

\begin{figure}
\caption{\label{slope}} Removal of the first electron vs.\ the peak intensity of a ten cycle 800\,nm sin$^2$-pulse (solid curve). A multiphoton-like behavior $\sim I^n$ with $n=5$ is evident. Instead, $n=13$ photons would be necessary to excite a 20\,eV Mie-resonance. Fragmentation threshold and C$_{60}^+$ saturation intensity from \cite{tchap} are also indicated (vertical lines).   
\end{figure}

\begin{figure}
\caption{\label{panelafterpulse}} Dipole $d(t)$ [plots (a) and (c)] and its Fourier transformation [plots (b) and (d)] for a ten cycle 800\,nm sin$^2$-pulse of peak intensity $I=4.56\times 10^{12}$\,W/cm$^2$ [plots (a) and (b)] and $I=2.85\times 10^{13}$\,W/cm$^2$ [plots (c) and (d)]. The black spectra are calculated from the entire time interval, the gray curves (multiplied by $10^4$, for better legibility) are spectra calculated from the time after the laser pulse. The former exhibit laser harmonics while in the latter single particle hole-excitations at low energies are dominant. Excitations $>15$\,eV are many orders of magnitude too weak for having a strong effect on ionization.
\end{figure}

\pagebreak

\thispagestyle{empty}

{\small
{\halign{ \strut #   &  \hfil #    &   \hfil #   &   \hfil #   &   \hfil #   &   \hfil #   &   \hfil #   &   \hfil #   &  \hfil #    &  \hfil #    &  \hfil #   &  \hfil #    \cr \noalign{\hrule}
$\ell=$& $0 $ &$1$ &$2$ &$3$ &$4$ &$5$ & $6$&$7$ &$8$ &$9$ &$10$  \cr \hline\hline
$n=1$ ($\sigma$) &-1.275 & -1.252 & -1.205 & -1.136 & -1.044 & -0.932 & -0.799 & -0.648 & -0.478 & -0.293 & {\em -0.092}  \cr
orb.\ no. & 0 & 1,2 & 3--5 & 6--9 & 10--14 & 15--20 & 21--27 & 28--35 & 36--44 & 45--54 \cr \noalign{\hrule}
$n=2$  ($\pi$)& -0.523 & -0.497 & -0.445 & -0.370 & -0.274 & {\em -0.159} & {\em -0.031} & {\em -0.006} & {\em -0.002} & &  \cr
orb.\ no. & 55 & 56,57 & 58--60 & 61--64 & 65--69  \cr \noalign{\hrule}
$n=3$ ($\delta$) & {\em -0.080} & {\em -0.068} & {\em -0.052} & {\em -0.034} & {\em -0.021} & {\em -0.014} & {\em -0.010} & & & &  \cr \noalign{\hrule}} }
}

\vfill

{\bf Table 1: D.~Bauer {\em et al.}, ``C$_{60}$ in intense femtosecond laser pulses ...''}

\pagebreak
\thispagestyle{empty}

\unitlength1cm
\begin{picture}(12,22)
\put(-2,0){\bild{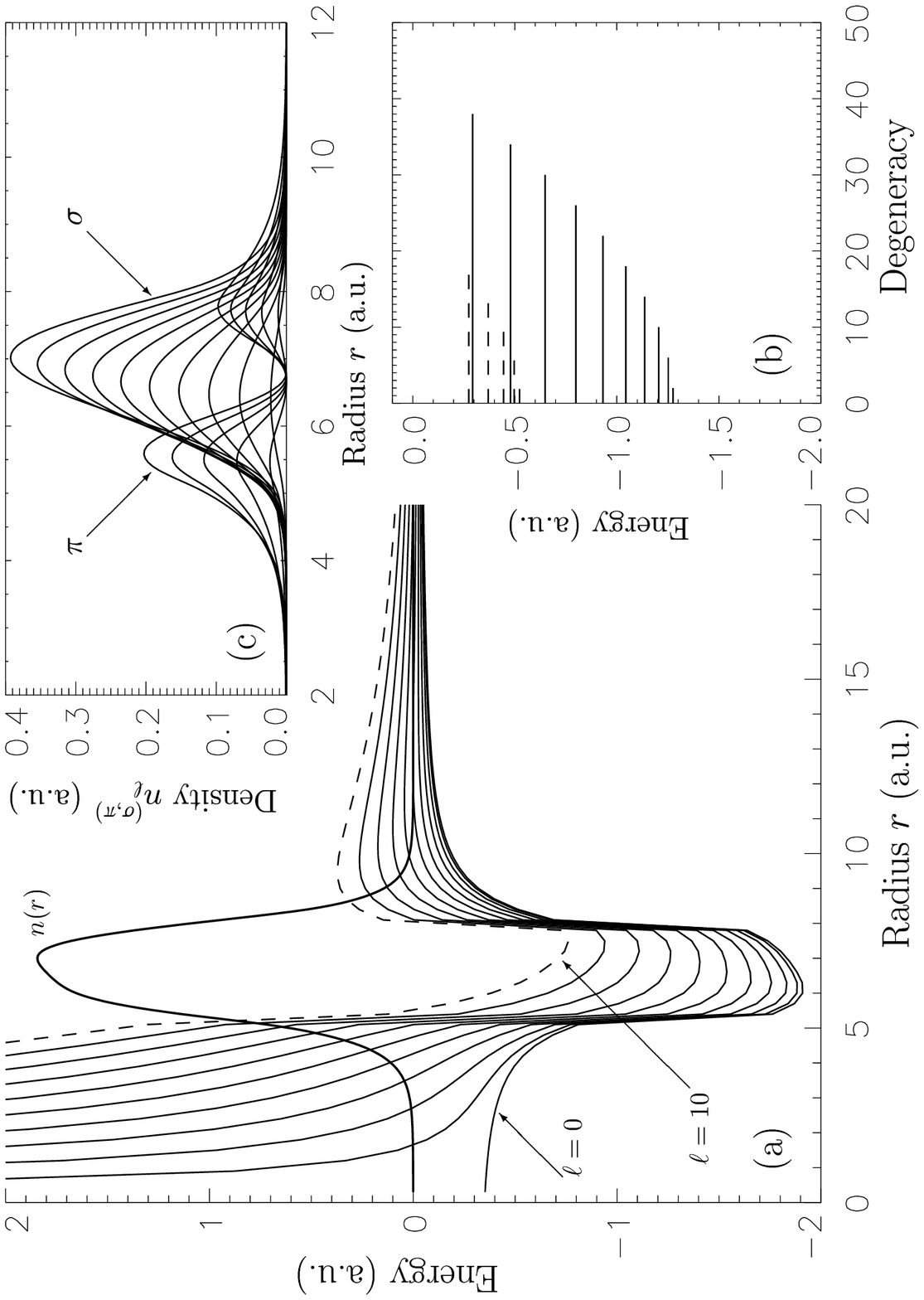}{18cm}}
\end{picture}

\vfill

{\bf Fig. 1: D.~Bauer {\em et al.}, ``C$_{60}$ in intense femtosecond laser pulses ...''}

\pagebreak
\thispagestyle{empty}

\unitlength1cm
\begin{picture}(12,12)
\put(0,0){\bild{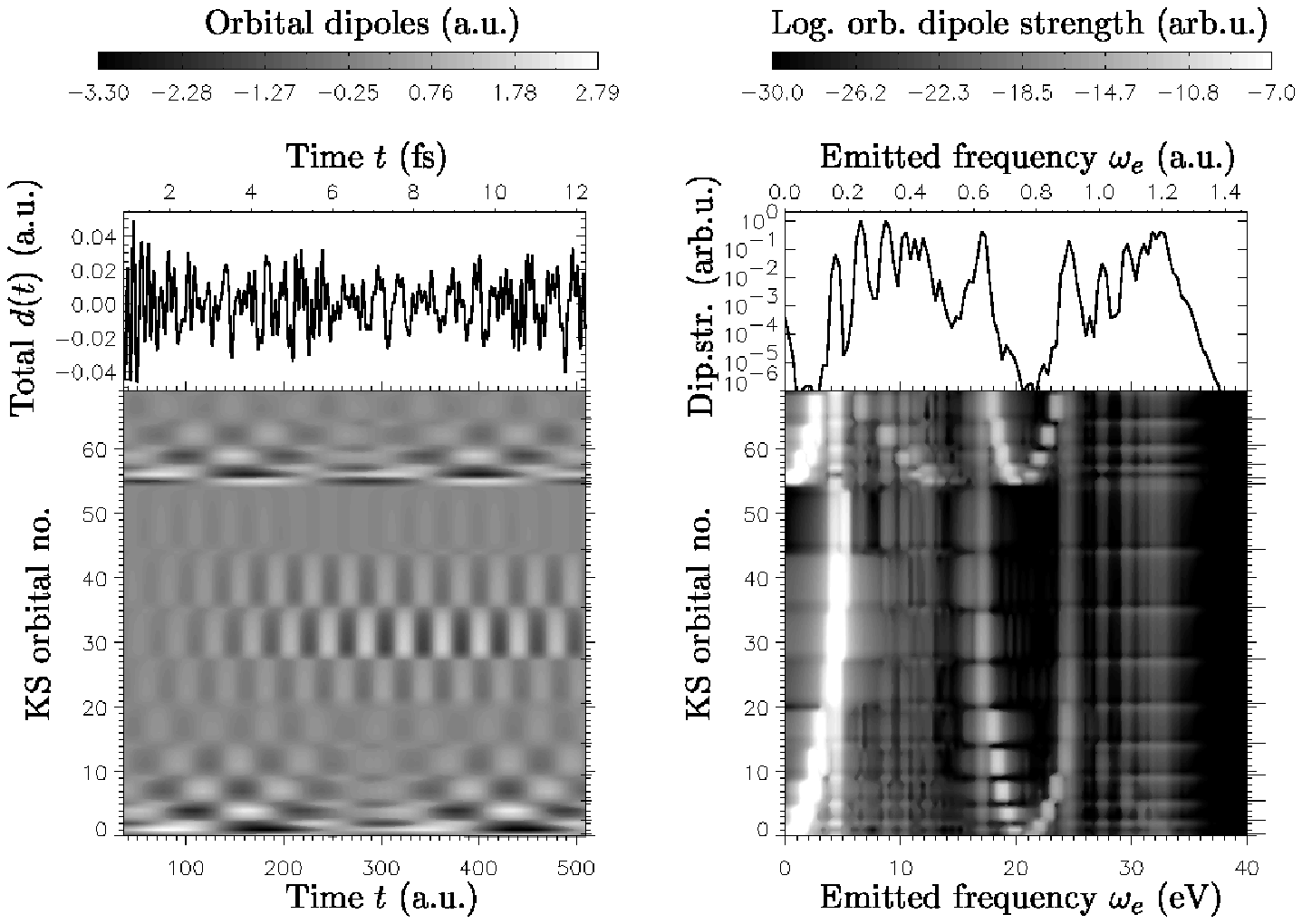}{16cm}}
\end{picture}

\vfill

{\bf Fig. 2: D.~Bauer {\em et al.}, ``C$_{60}$ in intense femtosecond laser pulses ...''}

\pagebreak
\thispagestyle{empty}

\unitlength1cm
\begin{picture}(12,12)
\put(0,0){\bild{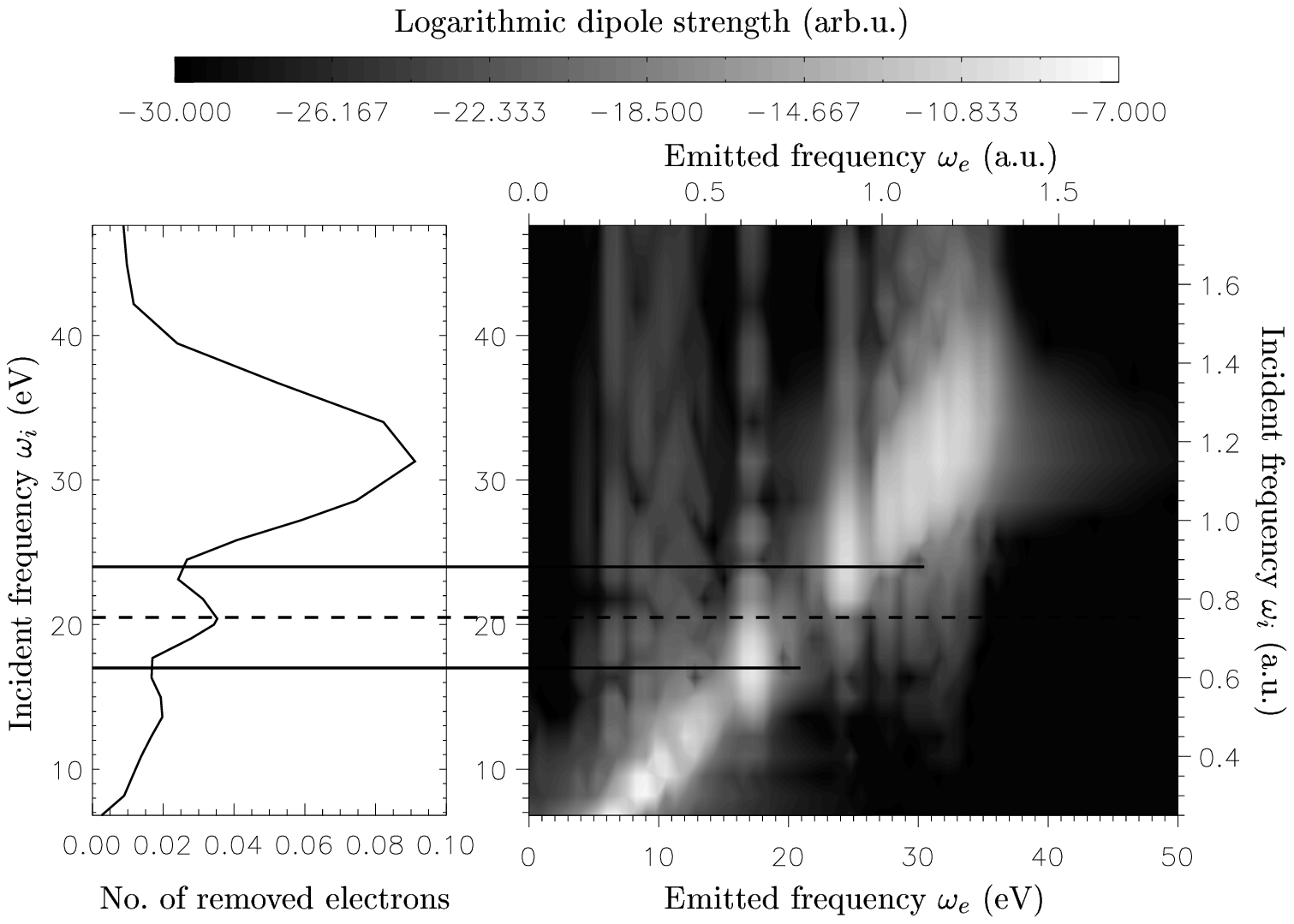}{20cm}}
\end{picture}

\vfill

{\bf Fig. 3: D.~Bauer {\em et al.}, ``C$_{60}$ in intense femtosecond laser pulses ...''}

\pagebreak
\thispagestyle{empty}

\unitlength1cm
\begin{picture}(12,13)
\put(0,0){\bild{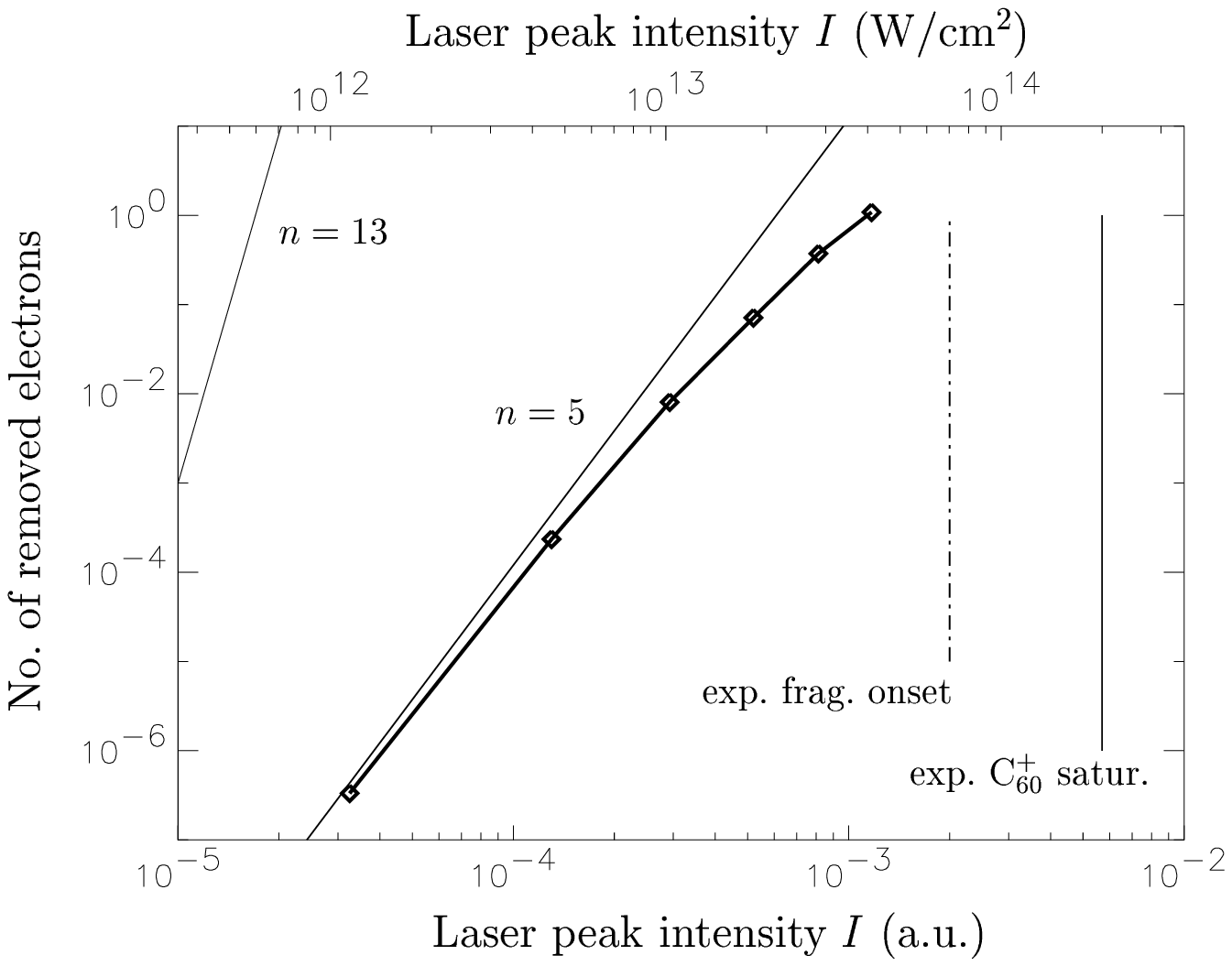}{14cm}}
\end{picture}

\vfill

{\bf Fig. 4: D.~Bauer {\em et al.}, ``C$_{60}$ in intense femtosecond laser pulses ...''}

\pagebreak
\thispagestyle{empty}

\unitlength1cm
\begin{picture}(12,13)
\put(0,0){\bild{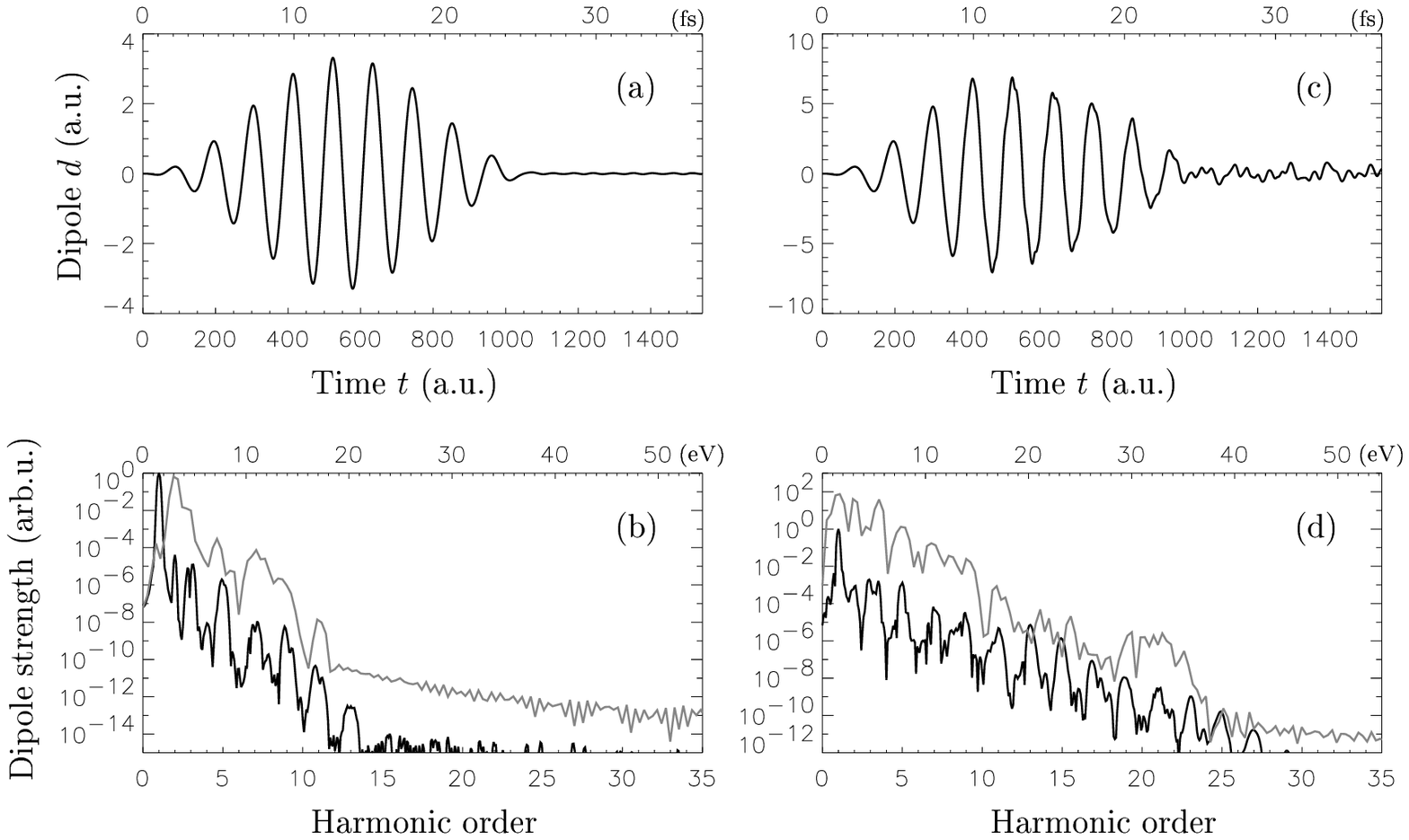}{20cm}}
\end{picture}

\vfill

{\bf Fig. 5: D.~Bauer {\em et al.}, ``C$_{60}$ in intense femtosecond laser pulses ...''}


\begin{thebibliography}{99}
\bibitem{revs} M.\ Protopapas, C.\ H.\ Keitel, and P.\ L.\ Knight, Rep.\ Progr.\ Phys.\ {\bf 60}, 389 (1997); C.\ J.\ Joachain, M.\ D\"orr, and N.\ J.\ Kylstra, Adv.\ At.\ Mol.\ Opt.\ Phys.\ {\bf 42}, 225 (2000).
\bibitem{taylor} E.\ S.\ Smyth, J.\ S.\ Parker, and K.\ T.\ Taylor, Comput.\ Phys.\ Commun.\ {\bf 114}, 1 (1998); J.\ S.\ Parker, L.\ R.\ Moore, D.\ D.\ Dundas, and K.\ T.\ Taylor, J.\ Phys.\ B: At.\ Mol.\ Opt.\ Phys.\ {\bf 33}, L691 (2000).
\bibitem{dreizgross} R.\ M.\ Dreizler and E.\ K.\ U.\ Gross, ``Density Functional Theory: An Approach to the Quantum Many-Body Problem,'' (Springer, Berlin, 1990).
\bibitem{runge} Erich Runge and E.\ K.\ U.\ Gross, Phys.\ Rev.\ Lett.\ {\bf 52}, 997 (1984).
\bibitem{burke} Kieron Burke and E.\ K.\ U.\ Gross, {\em A Guided Tour of Time-Dependent Density Functional Theory} in: ``Density Functionals: Theory and Applications'' ed.\ by Daniel Joubert, (Springer, Berlin, 1998), p.~116.
\bibitem{metclus} F.\ Calvayrac, P.-G.\ Reinhard, E.\ Suraud, and C.\ A.\ Ullrich, Phys.\ Rep.\ {\bf 337}, 493 (2000).
\bibitem{tddftnsi}  C.\ A.\ Ullrich and E.\ K.\ U.\ Gross, Comm.\ At.\ Mol.\ Phys.\ {\bf 33}, 211 (1997);  M.\ Petersilka and E.\ K.\ U.\ Gross, Laser Physics {\bf 9}, 105 (1999).
\bibitem{baueroe}  D.\ Bauer and F.\ Ceccherini, Optics Express {\bf 8}, 377 (2001).
\bibitem{campb} E.\ E.\ B.\ Campbell, K.\ Hansen, K.\ Hoffmann, G.\ Korn, M.\
  Tchaplyguine, M.\ Wittmann, and I.\ V.\ Hertel, Phys.\ Rev.\ Lett.\ {\bf
    84}, 2128 (2000).
\bibitem{tchap} M.\ Tchaplyguine, K.\ Hoffmann, O.\ D\"uhr, H.\ Hohmann, G.\
  Korn, H.\ Rottke, M.\ Wittmann, I.\ V.\ Hertel, and E.\ E.\ B.\ Campbell,
  J.\ Chem.\ Phys.\ {\bf 112}, 2781 (2000). 
\bibitem{hunsche} S.\ Hunsche, T.\ Starczewski, A.\ l'Huillier, A.\ Persson, C.-G.\ Wahlstr\"om, B.\ van Linden van den Heuvell, and S.\ Svanberg, Phys.\ Rev.\ Lett.\ {\bf 77}, 1966 (1996).
\bibitem{bertsch} George F.\ Bertsch, Aurel Bulgac, David Tom{\`a}nek, and Yang Wang, Phys.\ Rev.\ Lett.\ {\bf 67}, 2690 (1991).
\bibitem{hertel} I.\ V.\ Hertel, H.\ Steger, J.\ de Vries, B.\ Weisser, C.\ Menzel, B.\ Kamke, and W.\ Kamke, Phys.\ Rev.\ Lett.\ {\bf 68}, 784 (1992).
\bibitem{puska} M.\ J.\ Puska, and R.\ M.\ Nieminen, Phys.\ Rev.\ A {\bf 47},
  1181 (1993).
\bibitem{brack} M.\ Brack, Rev.\ Mod.\ Phys.\ {\bf 65}, 677 (1993).
\bibitem{ullr} C.\ A.\ Ullrich, P.-G.\ Reinhard, and E.\ Suraud, J.\ Phys.\ B: At.\ Mol.\ Opt.\ Phys.\ {\bf 31}, 1871 (1998).
\bibitem{perdwang} John P.\ Perdew and Yue Wang, Phys.\ Rev.\ B {\bf 45}, 13244 (1992). 
\bibitem{vshiv} V.\ A.\ Vshivkov, N.\ M.\ Naumova, F.\  Pegoraro, and S.\ V.\ Bulanov, Phys.\ Plasmas {\bf 5}, 2727 (1998).
\bibitem{explaindipole}  Taking into account the quadrupole also leads only to slight quantitative changes.
\bibitem{muller} H.\ G.\ Muller, Laser Physics {\bf 9}, 138 (1999).
\bibitem{faisalbook} F.\ H.\ M.\ Faisal, {\em Theory of Multiphoton Processes} (Plenum Press, New York, 1987).
\bibitem{seifert} G.\ Seifert, K.\ Vietze, and R.\ Schmidt, J.\ Phys.\ B: At.\ Mol.\ Opt.\ Phys.\ {\bf 29}, 5183 (1996), and references therein.
\bibitem{yabana} K.\ Yabana and G.\ F.\ Bertsch, Phys.\ Rev.\ B {\bf 54},  4484 (1996).
\bibitem{alasia} F.\ Alasia, R.\ A.\ Broglia, H.\ E.\ Roman, L.\ Serra, G.\ Colo, J.\ M.\ Pacheco,  J.\ Phys.\ B: At.\ Mol.\ Opt.\ Phys.\ {\bf 27}, L643 (1994).
\bibitem{ullrii} C.\ A.\ Ullrich, P.-G.\ Reinhard, and E.\ Suraud, J.\ Phys.\ B: At.\ Mol.\ Opt.\ Phys.\ {\bf 30}, 5043 (1997).
\bibitem{bauerati} D.\ Bauer {\em et al.}, unpublished.
\end{thebibliography}
\end{document}